\documentclass[12pt]{article}
\usepackage{amsmath,amsfonts,amssymb}
\usepackage{geometry}
\usepackage{natbib}
\usepackage{graphicx}
\usepackage[colorlinks=true,linkcolor=red,citecolor=blue]{hyperref}
\title{The efficiencies of pilot feasibility trials in rare diseases using Bayesian methods}
\author
{\small \textbf{Lara Maleyeff}$^{1,*}$, \textbf{Shirin Golchi}$^{1}$, \textbf{Val\'erie Leclair}$^{2,3}$, \textbf{and }\textbf{Marie Hudson}$^{2,3}$\\
\footnotesize $^{1}$Department of Epidemiology, Biostatistics, and Occupational Health, McGill University, \\  
\footnotesize  
2001 McGill College Avenue, Montr\'eal, QC, H3A 1Y7, CA \\
\footnotesize $^{2}$Department of Medicine, McGill University, \\
\footnotesize  
3605 Rue de la Montagne, Montr\'eal, QC, H3G 2M1, CA \\
\footnotesize $^{3}$Jewish General Hospital and Lady Davis Institute for Medical Research,\\
\footnotesize 3755 Chemin de la Côte-Sainte-Catherine, Montr\'eal, QC, H3T 1E1, CA\\
\bigskip
\footnotesize \textsuperscript{*}\textit{email}: lara.maleyeff@mcgill.ca
}

\date{}

\begin{document}

\maketitle
\newpage
\begin{abstract}
\noindent
Pilot feasibility studies play a pivotal role in the development of clinical trials for rare diseases, where small populations and slow recruitment often threaten trial viability. While such studies are commonly used to assess operational parameters, they also offer a valuable opportunity to inform the design and analysis of subsequent definitive trials—particularly through the use of Bayesian methods. In this paper, we demonstrate how data from a single, protocol-aligned pilot study can be incorporated into a definitive trial using robust meta-analytic-predictive priors. We focus on the case of a binary efficacy outcome, motivated by a feasibility trial of intravenous immunoglobulin tapering in autoimmune inflammatory myopathies. Through simulation studies, we evaluate the operating characteristics of trials informed by pilot data, including sample size, expected trial duration, and the probability of meeting recruitment targets. Our findings highlight the operational and ethical advantages of leveraging pilot data via robust Bayesian priors, and offer practical guidance for their application in rare disease settings.

\end{abstract}

\section{Introduction}
Rare diseases present unique challenges for clinical research. According to \citet{griggs2009clinical}, rare diseases often affect fewer than 200,000 individuals in the United States, necessitating innovative clinical trial designs due to limited patient populations. Recruitment difficulties are compounded by the need for geographically dispersed collaborations and specialized expertise. In these settings, feasibility studies—or pilot trials—play a crucial role in ensuring that definitive clinical trials are both logistically and scientifically sound. These preliminary studies help identify and address key uncertainties in trial conduct, such as recruitment and retention rates, protocol adherence, data quality, and outcome measurement procedures \citep{arain2010, thabane2010guidelines}. Given the high logistical and financial burden of confirmatory clinical trials in rare diseases, conducting well-designed feasibility studies is a critical step toward ensuring trial success.

Despite their importance, pilot trials are often treated solely as stand-alone assessments and not leveraged beyond initial feasibility evaluation. However, when designed with sufficient rigor and aligned closely with the planned definitive trial—in terms of eligibility criteria, outcome definitions, protocols, and data collection methods—feasibility studies offer an opportunity to generate high-quality data that can inform both design decisions and statistical inference in the subsequent trial. Doing so not only improves efficiency but may also address ethical concerns associated with data waste, particularly in the context of rare diseases where every patient outcome contributes significantly to the body of knowledge about the disease. In this way, feasibility studies can serve a dual purpose: identifying operational barriers and providing a reliable source of prior information for formal incorporation into the analysis of the definitive trial.

Bayesian methods offer a natural and flexible framework for incorporating prior information. While much of the existing literature focuses on the use of historical data, the same methods apply when the prior data are drawn from an earlier feasibility study that was deliberately harmonized with the main trial. Regulatory guidance from the U.S. Food and Drug Administration (FDA) supports this approach, emphasizing that prior information can be used when assumptions of exchangeability and compatibility are adequately justified \citep{fda2010bayesian}. Borrowing from feasibility data under these conditions has the potential to reduce required sample size, improve trial efficiency, and increase the probability of trial success without compromising statistical validity.

Although feasibility studies can produce high-quality data, directly pooling pilot data with definitive trial data is generally discouraged due to unavoidable protocol modifications and the potential for unrecognized sources of variability. Even minimal changes between pilot and main trial designs can introduce additional variation and compromise validity of a pooled analysis \citep{leon2011pilot}. A more principled alternative is to incorporate pilot data through informative priors, which allow for calibrated borrowing based on study compatibility and reduce the risk of bias in confirmatory analyses.

Several Bayesian borrowing methods have been proposed to mitigate potential biases arising form heterogeneity between data sources. The power prior \citep{ibrahim2000power} incorporates historical data by raising its likelihood to a scaling factor, allowing for calibrated borrowing. The normalized power prior \citep{duan2006evaluating} uses a scaling factor to control the influence of historical data, while the commensurate prior \citep{hobbs2011hierarchical} incorporates explicit measures of similarity to modulate borrowing. Supervised priors \citep{pan2017calibrated} rely on predefined compatibility metrics to adjust informativeness manually. The robust meta-analytic-predictive (MAP) prior \citep{schmidli2014robust} offers a principled compromise between efficiency and robustness. It combines a weakly informative component with a prior derived from historical data, mitigating the risk of prior-data conflict while still allowing meaningful information borrowing. This property is particularly valuable in early stage studies, where full prior-data agreement cannot be guaranteed. \citet{qi2022sample} extended this framework to support Bayesian sample size determination, demonstrating that MAP priors can lead to substantial reductions in required sample size when historical and current data are commensurate—without inflating type I error rates. While other methods may be equally or more efficient under ideal conditions, simulation studies by \citet{psioda2019bayesian} suggest that under stringent type I error constraints, simpler models such as partial-borrowing power priors often perform comparably. In light of these trade-offs, we selected the robust MAP prior for this study due to its balance of flexibility, interpretability, and its demonstrated ability to preserve statistical validity even in the presence of modest prior-data conflict.

Although robust MAP priors are designed to mitigate the effects of prior-data conflict, their performance improves substantially when the pilot and definitive trials are highly compatible. If important differences exist between the studies—such as in patient populations, intervention protocols, or outcome definitions—the relevance of pilot data diminishes, and the potential benefits of borrowing are reduced. To preserve the value of pilot data and maximize its contribution as a prior, it is essential to consider compatibility at the pilot study design stage. Eligibility criteria, endpoints, follow-up procedures, and operational settings must be harmonized as much as possible from the outset to ensure data exchangeability and justify the use of MAP priors.

Furthermore, defining the confirmatory hypothesis during the design of the pilot study helps avoid the risk of hypothesizing after results are known (HARKing). This practice, often driven by exploratory findings and retrospective adaptations, compromises scientific validity by inflating type I error and yielding overly optimistic inferences. As emphasized by \citet{fleming2010clinical}, post hoc hypothesis revisions undermine the integrity of confirmatory research and increase the likelihood of false discoveries. By prospectively specifying the confirmatory hypothesis and embedding it in both the pilot and definitive trial designs, investigators preserve the objectivity of the analysis and ensure that the pilot contributes meaningfully not only to feasibility assessment, but also to the rigor and credibility of the overall trial program.

The present paper illustrates the practical benefits of using pilot study data as informative priors through the commonly used robust MAP approach. Specifically, we demonstrate how this strategy can substantially reduce the required sample size, shorten expected trial duration, and increase the probability of reaching recruitment targets under realistic accrual scenarios. These approaches, originally developed for integrating historical control data, are especially well suited for rare disease trials, where small sample sizes and high uncertainty make efficient and ethical designs essential. Section~\ref{sec:trial} introduces the IVIg de-escalation feasibility trial in autoimmune inflammatory myopathies, which motivates our methodological focus on using pilot data to inform prior distributions for a subsequent definitive trial. In Section~\ref{sec:map}, we present the statistical framework for the confirmatory analysis, illustrating how robust MAP priors can be constructed from pilot data in both arms to enable partial borrowing while mitigating the risks of prior-data conflict. In Section~\ref{sec:sim}, we present a simulation study to evaluate how the size of the pilot study impacts power, required sample size, and trial duration under different control event risks and treatment effects. Finally, Section~\ref{sec:discussion} discusses the implications of our findings for rare disease trial design and highlights opportunities for future research.

\section{IVIg De-Escalation Feasibility Trial}
\label{sec:trial}

Autoimmune inflammatory myopathies (AIM) are rare chronic diseases that cause muscle weakness and inflammation. Intravenous immunoglobulin (IVIg) is widely used to treat AIM, especially when other treatments are ineffective \citep{aggarwal2022trial, lim2021intravenous}. It has been shown to improve outcomes in patients with dermatomyositis, leading to its approval as the first and only FDA-approved treatment for this condition \citep{aggarwal2022trial}. However, there is little evidence to guide how long IVIg should be continued or how it should be tapered. This is concerning, as IVIg is costly, can cause serious side effects such as blood clots—particularly in AIM patients, who already face elevated thrombotic risk—and is susceptible to shortages \citep{inspq2023ivig}. 

The IVIg de-escalation feasibility trial, with a planned sample size of 30 patients, is designed to assess key feasibility outcomes such as recruitment rates, adherence to tapering regimens, and loss to follow-up. As part of the trial design, patients will be randomized to one of two de-escalation strategies: one guided by routine clinical lab tests (which can be poor predictors of disease flares), and the other guided by novel biomarkers—reflecting an underlying hypothesis that precision medicine approaches may ultimately be more effective than standard methods. Secondary safety outcomes will include the proportion of patients able to discontinue IVIg without experiencing a flare, time to flare, and stability of relevant biomarkers. These secondary efficacy data will be used to construct informative yet robust MAP priors for a subsequent confirmatory Phase III randomized controlled trial, allowing for calibrated incorporation of early-phase evidence.

For the definitive Phase III trial, the goal is to use feasibility data to inform both prior distributions and design parameters. A Beta prior will be derived from the observed proportion of IVIg discontinuation success in the feasibility study, and Bayesian sample size determination will be conducted to ensure adequate power. Assuming commensurability between the pilot and confirmatory trials, an estimated 180 patients per arm will be required to achieve 80\% power while maintaining type I error rate control.

\section{Statistical Methods}
We use the case of dichotomous outcomes—motivated by the IVIg de-escalation feasibility trial—to illustrate how pilot data can be incorporated into the design and analysis of a subsequent confirmatory trial. In this trial, the key exploratory efficacy outcome is the binary indicator of successful IVIg discontinuation without flare. While feasibility data are not typically powered for inference, they can be highly valuable for informing prior distributions and guiding Bayesian sample size determination in a subsequent trial.
\label{sec:map}
\subsection{Notation}

Let \( p_C \) and \( p_T \) denote the true probabilities of success in the control and treatment arms, respectively. The sample sizes for the control and treatment arms in the Phase III trial are denoted as \( n_C^{(2)} \) and \( n_T^{(2)} \), and the corresponding numbers of observed successes are \( y_C^{(2)} \) and \( y_T^{(2)} \). In the preceding feasibility study, we observe \( y_C^{(1)} \) successes out of \( n_C^{(1)} \) participants in the control group, and \( y_T^{(1)} \) out of \( n_T^{(1)} \) in the treatment group.
\subsection{Confirmatory Study Hypothesis}
The primary objective of the Phase III trial is to evaluate whether IVIg tapering using novel biomarkers leads to superior rates of discontinuation without flare compared to standard of care. To this end, we frame a one-sided Bayesian hypothesis test:
\begin{equation}
    H_0: p_T \le p_C \quad \text{vs.} \quad H_1: p_T > p_C.
\end{equation}

\noindent In Bayesian clinical trials, decision are commonly made based on evaluating the posterior probability that the research hypothesis is true, i.e., \( P(p_T > p_C \mid y_C^{(2)}, y_T^{(2)}) \), and declaring treatment superiority if the posterior probability exceeds a threshold \( \phi \), chosen to control the frequentist type I error rate.

\subsection{Incorporating Pilot Data Using Robust MAP Priors}

Pilot studies serve as an essential preliminary step in trial development by evaluating feasibility of recruitment, adherence, outcome assessment, and study procedures. However, pilot data are not typically included directly in the analysis of the subsequent definitive trial. As \cite{leon2011pilot} caution, even minor protocol modifications between pilot and main trials introduce potential sources of bias and additional variability that can invalidate pooled analyses. Moreover, combining pilot and main trial data without accounting for their separate origins may inflate type I error rates and overstate precision.

To address these issues while preserving the value of pilot data, we incorporate information from the pilot study using a robust MAP prior for \( p_C \) and \( p_T \) \citep{schmidli2014robust}. Although the MAP prior is formally defined via a Bayesian hierarchical model over multiple historical studies, our setting involves a single pilot study. Following \cite{schmidli2014robust}, we approximate the MAP using a mixture prior that includes an informative component based on the pilot data and a vague component. This avoids the need to specify a between-study heterogeneity parameter, which cannot be estimated with only one source of historical data. As such, our approach implicitly assumes a fixed, low level of heterogeneity, while the vague component and data-driven weight updating help guard against prior-data conflict. For each group \( g \in \{C, T\} \), we define a prior on \( p_g \) as a mixture of an informative component, derived from the pilot study, and a vague component:

\begin{equation}
    p_g \mid w, \alpha_g, \beta_g \sim (1 - w) \cdot \text{Beta}(1, 1) + w \cdot \text{Beta}(\alpha_g, \beta_g),
\end{equation}

\noindent where the informative prior parameters are computed via conjugate updating from the pilot data:
\begin{equation}
    \alpha_g = a_0 + y_g^{(1)}, \quad \beta_g = b_0 + n_g^{(1)} - y_g^{(1)},
\end{equation}

\noindent with \( a_0 = b_0 = 1 \) representing a weakly informative prior. We use a default initial prior weight \( w = 0.5 \), balancing robustness and informativeness. 

The likelihood for each group in the definitive trial follows from the model,
\begin{equation}
    y_g^{(2)} \mid n_g^{(2)}, p_g \sim \text{Binomial}(n_g^{(2)}, p_g),
\end{equation}

\noindent yielding a posterior distribution for \( p_g \) that is itself a weighted mixture of Beta distributions:

\begin{equation}
    p_g \mid y_g^{(2)} \sim (1 - \tilde{w}_g) \cdot \text{Beta}(1 + y_g^{(2)}, 1 + n_g^{(2)} - y_g^{(2)}) + \tilde{w}_g \cdot \text{Beta}(\alpha_g + y_g^{(2)}, \beta_g + n_g^{(2)} - y_g^{(2)}).
\end{equation}

\noindent The updated weight on the informative component, \( \tilde{w}_g \), is computed based on the marginal likelihood under each prior component:

\[
\tilde{w}_g = \frac{w \cdot f_{\text{inf}}}{w \cdot f_{\text{inf}} + (1 - w) \cdot f_{\text{vague}}},
\]

\noindent where \( f_{\text{inf}} \) and \( f_{\text{vague}} \) are the marginal likelihoods under the informative and vague priors, respectively. This approach allows partial borrowing that adapts based on consistency between pilot and current data \citep{schmidli2014robust}.

The robust MAP prior yields a posterior that remains a weighted mixture of conjugate distributions, such as Beta distributions for binary outcomes, enabling closed-form updates and straightforward uncertainty quantification. This mixture structure provides a principled way to downweight historical information in the presence of prior-data conflict, enhancing the robustness of the analysis. While we focus on binary outcomes in this work, the robust MAP framework is general and applies to any setting where conjugate priors exist within the exponential family (e.g., Gaussian for continuous data, Gamma for rates). It can also be extended to accommodate hierarchical models that integrate multiple historical sources and account for between-study heterogeneity \citep{schmidli2014robust}. In more complex scenarios—such as time-to-event data—non-conjugate formulations and MCMC-based estimation allow for flexible implementation of MAP priors, expanding the applicability of these methods to a broad array of clinical trial designs \citep{bi2023beats}.

\section{Simulation Study}
\label{sec:sim}
We conducted a simulation study to evaluate how incorporating pilot study data influences the required sample size and trial feasibility when using robust MAP priors. We considered a one-sided Bayesian hypothesis test to evaluate treatment superiority. The null hypothesis was that the event probability under the treatment arm is less than or equal to that of the control arm, \( H_0: p_T \le p_C \), versus the alternative \( H_1: p_T > p_C \). 

For each scenario, we considered three baseline control event probabilities: \( p_C \in \{0.06, 0.25, 0.6\} \). Treatment success probabilities were determined by multiplying the control rate by a corresponding risk ratio (RR), with values \( \text{RR} \in \{1.3, 1.7, 1.9\} \), yielding \( p_T = \text{RR} \cdot p_C \). These values were fixed across both the pilot and definitive trials, representing a setting with no prior-data conflict. For each combination of \( p_C \), RR, and pilot study proportion (0\%, 10\%, 20\%, 30\%, or 40\% of the definitive sample size), we performed 10{,}000 simulation replicates.

In each replicate, we first generated pilot study data by simulating binary outcomes from independent Bernoulli distributions using the true probabilities \( p_C \) and \( p_T \), with sample sizes determined by the specified pilot-to-definitive ratio with 1:1 allocation. The observed numbers of successes in the pilot arms were recorded. Next, for the given definitive trial sample size $n^{(2)} = n_C^{(2)} + n_T^{(2)} $, we again simulated binary outcomes for both the control and treatment arms using the same true probabilities with 1:1 allocation. We then computed the posterior probability that the treatment was superior to control using a robust MAP prior informed by the pilot data and a weakly informative component (see Section \ref{sec:map} for more detail). If the posterior probability \( P(p_T > p_C \mid Y_T^{(2)}, Y_C^{(2)}) \) exceeded 0.975, the trial was considered to have detected a significant treatment effect. For each setting, we identified the minimum value of \( n^{(2)} \) required to achieve 80\% power across simulations.

To evaluate the practical implications of our sample size recommendations, we computed the expected trial duration under varying recruitment rates. Specifically, we assumed monthly recruitment rates of 2, 5, and 10 participants per arm and calculated the expected trial duration analytically as \( \text{Duration} = n / \lambda \), where \( n \) is the required sample size (from the power analysis) and \( \lambda \) is the recruitment rate. For each combination of pilot data proportion, baseline control risk, and recruitment rate, we computed the corresponding trial duration and examined how the incorporation of external data (via the pilot study) could reduce trial timelines.

\begin{figure}[ht]
\centering
\includegraphics[width=\textwidth]{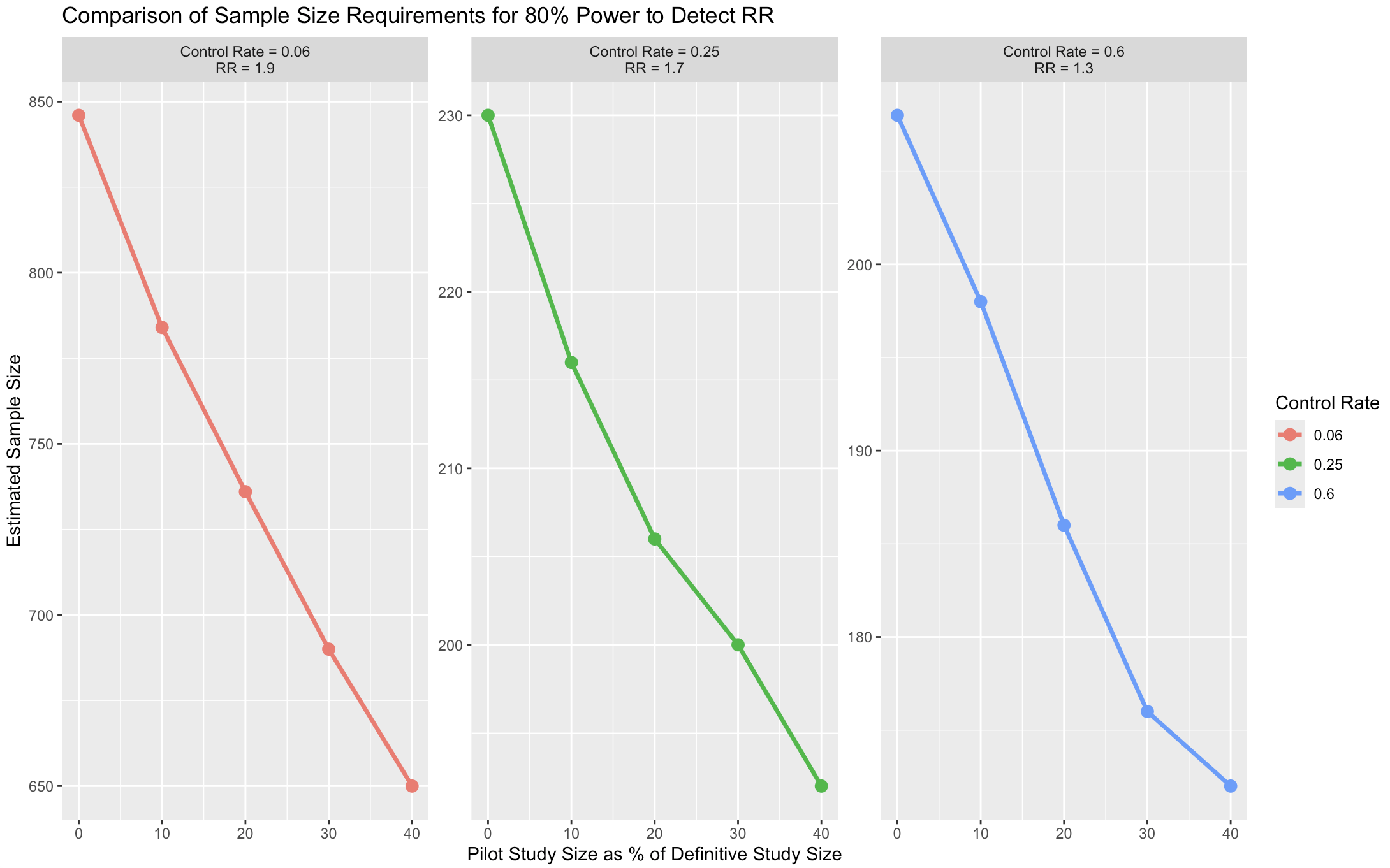}
\caption{Required total sample size to achieve 80\% power across different pilot study sizes, baseline control risks, and risk ratios.}
\label{fig:sample_size}
\end{figure}

In addition, we estimated the probability of successfully recruiting the required sample size within a given trial duration using methods similar to those used by \cite{churipuy2024bayesian} and originally proposed by \cite{anisimov2007modelling}. Let \( \lambda_0 \) represent the average recruitment rate per month per arm observed in the pilot study. To account for uncertainty in this rate, we assigned a Gamma prior to the true site-level recruitment rate:
\begin{equation}
    \lambda \mid \lambda_0 \sim \text{Gamma}(2\lambda_0, 2),
\end{equation}
\noindent which reflects moderate confidence around the pilot estimate, with mean \( \lambda_0 \) and variance \( \lambda_0 / 2 \). Assuming a Poisson recruitment process over \( m \) months, the total number of recruits \( N \) in one arm follows:
\begin{equation}
    N \mid \lambda, m \sim \text{Poisson}(\lambda m).
\end{equation}

\noindent Marginalizing over the Gamma prior yields a Negative Binomial distribution:
\begin{equation}
    N \mid \lambda_0, m \sim \text{NegBin}(r = 2\lambda_0, p = \frac{2}{2 + m}),
\end{equation}
where \( r \) is the shape of the Gamma prior and \( p \) is derived from the rate parameter and trial duration. This allows us to compute:
\begin{equation}
    P(N \geq n \mid \lambda_0, m),
\end{equation}

\noindent the probability that the number of recruits exceeds the required sample size \( n \) within \( m \) months. We computed this probability for each scenario (combination of pilot size, control risk, and recruitment rate), allowing us to quantify how incorporating pilot data affects not only statistical power but also the likelihood of meeting recruitment goals in a timely manner.

To evaluate the impact of prior-data conflict, we conducted a secondary set of simulations in which the treatment effect observed in the pilot study differed from that of the definitive trial. Specifically, for a pilot study comprising 20\% of the definitive trial sample size, we considered scenarios in which the risk ratio in the pilot was smaller than in the definitive trial, with values equal to $\{0.80, 0.85, 0.90, 0.95\} \times \text{RR}$. This allowed us to assess how increasingly pessimistic pilot results affect the posterior distribution under robust MAP priors and, in turn, influence the required sample size to achieve 80\% power in the confirmatory trial.

All simulations were conducted in \texttt{R} (version 4.2.4) and the code is provided in the Supplementary Materials. While we implemented all methods independently, several R packages are available to support Bayesian clinical trial design using historical data. The \texttt{RBesT} package \citep{Weber2021} implements MAP priors, allowing users to synthesize historical control information and compute prior effective sample sizes. The \texttt{bayesDP} package \citep{Balcome2021} implements the power prior using a data-dependent discount function based on prior-data conflict; it is typically used alongside the \texttt{bayesCT} package \citep{Chandereng2020} for trial simulation and design. The \texttt{BayesCTDesign} package \citep{Eggleston2019} supports power prior-based Bayesian two-arm trial design using historical control data, although it does not allow for covariates or random power parameters. The \texttt{BayesPPD} package \citep{Shen2021arxiv} offers a comprehensive and flexible implementation of the power and normalized power prior for both model fitting and sample size determination across a wide range of outcome distributions, including support for generalized linear models with covariates.

\subsection{Results}
Incorporating pilot data consistently reduced the required sample size across all scenarios (Figure~\ref{fig:sample_size}). When the control event rate was low (\( p_C = 0.06 \)) and the treatment effect was large (\( \text{RR} = 1.9 \); left panel), including a pilot study equal to 20\% of the total required sample size of 736 (i.e., 147 participants) reduced the required number of new enrollees from 846 to 736---a 13\% reduction. Increasing the pilot contribution to 40\% of the final total sample size of 650 (i.e., 260 participants) yielded a 23\% reduction. For a moderate control event rate (\( p_C = 0.25 \); middle panel), including 20\% pilot data (41 of 206) reduced the total sample size from 230 to 206---a 10\% reduction. With 40\% pilot data (77 of 192), the sample size decreased further to 192---a 17\% reduction. Similarly, for a high control event rate (\( p_C = 0.6 \); right panel), including 20\% pilot data (37 of 186) reduced the sample size from 208 to 186 (an 11\% reduction), and 40\% pilot data (69 of 172) further reduced it to 172 (a 17\% reduction).

In the IVIg de-escalation trial, the anticipated flare rate in the standard-of-care arm is 0.25. Our simulation study assumes a one-sided alternative where $p_T > p_C$, expressed as $RR > 1$, although the framework is general and can be reparameterized for settings where a reduction in the event rate is desirable (i.e., $p_T < p_C$), as in this motivating example. With a pilot study comprising 20\% of the definitive study sample size, the required sample size for the definitive trial would decrease from 230 to 206 to detect a RR of 0.59 ($p_T = 0.15$), and from 1,402 to 1,244 to detect a RR of 0.75 ($p_T = 0.19$).
\begin{figure}[ht]
\centering
\includegraphics[width=\textwidth]{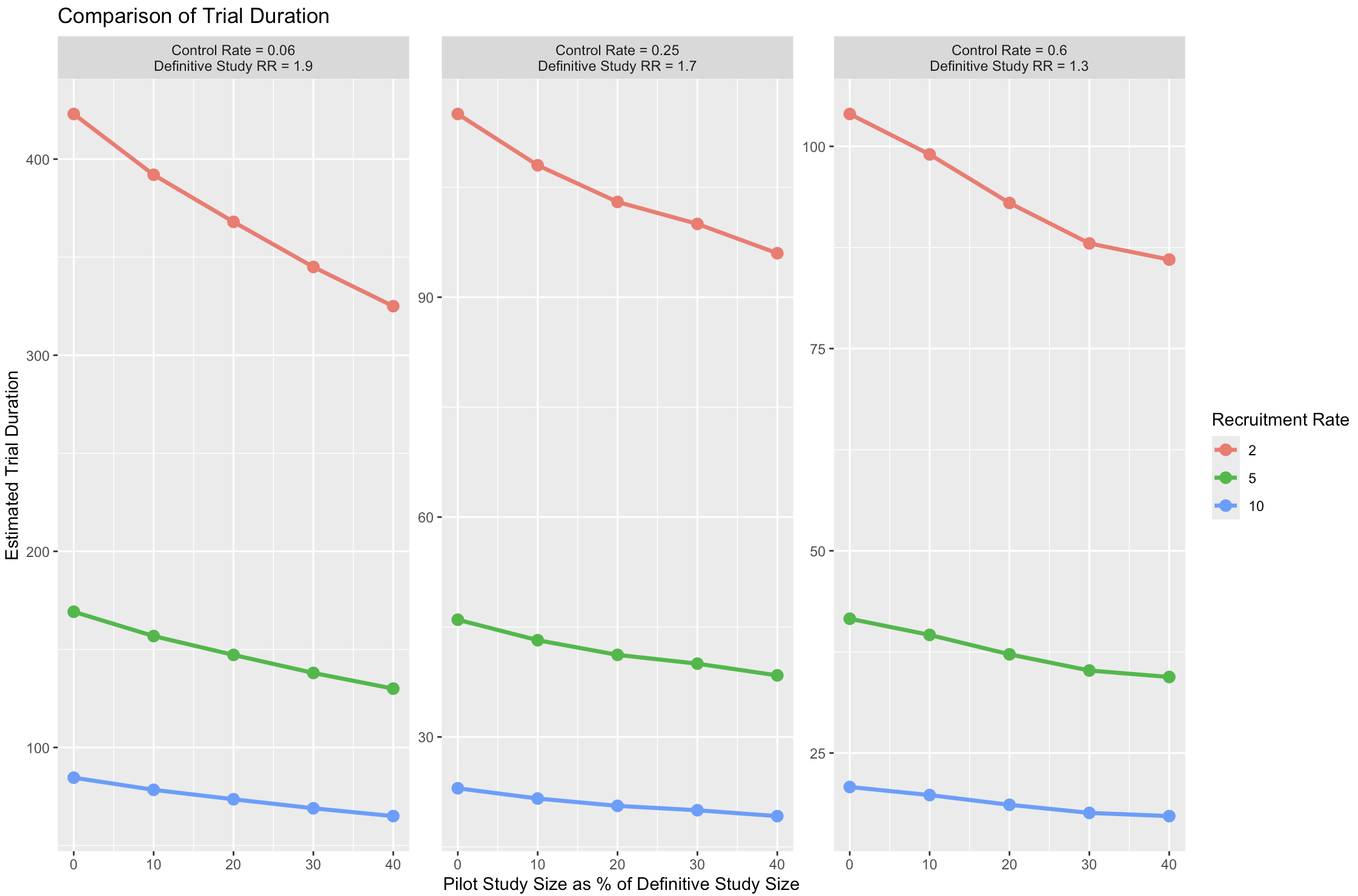}
\caption{Expected trial duration (in months) as a function of pilot data proportion and recruitment rate. Results are shown for three baseline control event probabilities.}
\label{fig:trial_duration}
\end{figure}

These reductions resulted in shorter expected trial durations under realistic recruitment rates (Figure~\ref{fig:trial_duration}). Across all control risks and recruitment rates (2, 5, or 10 participants per arm per month), increasing the size of the pilot data consistently shortened the expected durations of the trial. At a recruitment rate of 10 patients per month, increasing pilot data from 0\% to 20\% reduced the expected duration from 85 to 74 months when \( p_C = 0.06 \), from 23 to 21 months when \( p_C = 0.25 \), and from 21 to 19 months when \( p_C = 0.6 \). At 5/month, durations dropped from 169 to 147 months (\( p_C = 0.06 \)), from 46 to 41 months (\( p_C = 0.25 \)), and from 42 to 37 months (\( p_C = 0.6 \)). For the slowest recruitment rate of 2/month, durations decreased from 115 to 103 months at \( p_C = 0.25 \) and from 104 to 93 months at \( p_C = 0.6 \). The scenario with \( p_C = 0.06 \) and recruitment of 2/month, requiring over 35 years to complete, was deemed infeasible and is not interpreted further. 

The magnitude of duration reduction was greatest in absolute terms when recruitment was slower and required sample sizes were larger, as in the low control risk setting. Nevertheless, pilot data contributed meaningfully to trial efficiency across all plausible recruitment scenarios. These time savings are especially important in rare disease contexts, where recruitment is often slow and delays in evidence generation can limit timely access to effective therapies. Moreover, anticipated improvements in trial duration can inform the number of sites needed to meet recruitment targets within a desired timeframe, further aiding in trial planning and resource allocation.

\begin{figure}[ht]
\centering
\includegraphics[width=\textwidth]{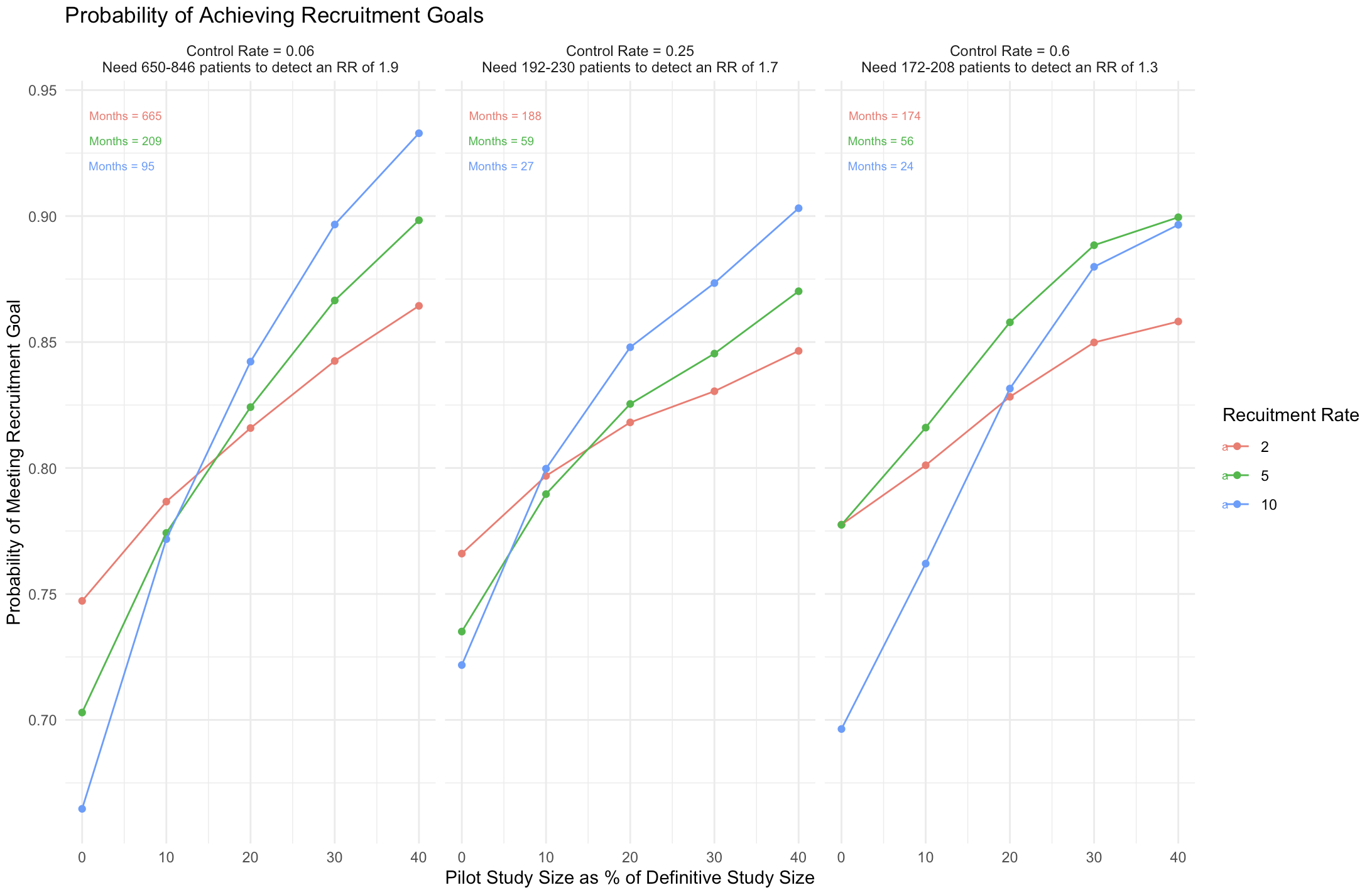}
\caption{Probability of meeting recruitment targets within a fixed trial duration, using a Gamma-Poisson model.}
\label{fig:recruitment_prob}
\end{figure}

Figure~\ref{fig:recruitment_prob} shows that incorporating pilot data consistently increased the probability of meeting recruitment targets within a fixed duration across all control risks and recruitment rates. When \(p_C = 0.06\), the probability of meeting recruitment goals increased from 0.70 to 0.82 at 5/month and from 0.67 to 0.84 at 10/month with 0\% vs. 20\% pilot data. For \(p_C = 0.25\), gains ranged from 0.77 to 0.82 (rate = 2), 0.74 to 0.83 (rate = 5), and 0.72 to 0.85 (rate = 10). Even with higher baseline probabilities at \(p_C = 0.6\), improvements remained notable: 0.78 to 0.83 (rate = 2), 0.78 to 0.86 (rate = 5), and 0.70 to 0.83 (rate = 10). These results highlight that modest pilot studies can meaningfully improve the feasibility of completing recruitment on time, even in high-accrual settings.

\begin{figure}[htbp]
\centering
\includegraphics[width=\textwidth]{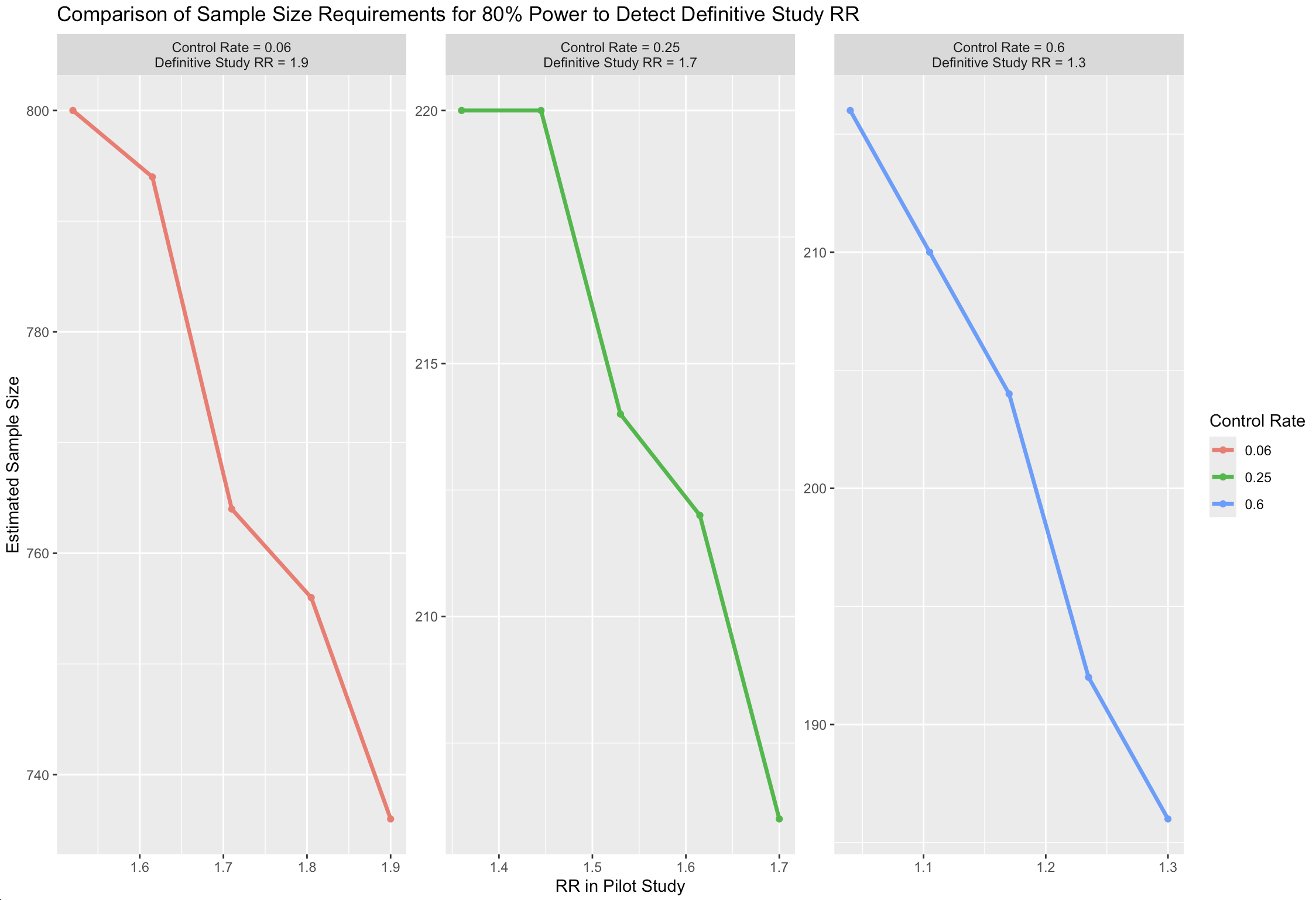}
\caption{Estimated definitive trial sample sizes required for 80\% power under varying levels of prior-data conflict. The right-most point on each plot represents the case with no prior-data conflict. For comparison, the required sample sizes when no pilot data are used are shown in Figure~\ref{fig:sample_size}.}
\label{fig:conflict}
\end{figure}

Figure~\ref{fig:conflict} displays the estimated definitive study sample sizes required to achieve 80\% power when incorporating pilot data with varying degrees of prior-data conflict. For each scenario, the pilot study sample size was fixed at 20\% of the definitive trial, and the risk ratio in the pilot study ranged from 80\% to 95\% of the true risk ratio used in the definitive trial. When the control event probability was $p_C = 0.06$ or $p_C = 0.25$, the required sample size consistently decreased as the degree of agreement between the pilot and definitive trials improved. Importantly, even under the most extreme prior-data conflict ($\text{RR}_{\text{pilot}} = 0.8 \times \text{RR}_{\text{true}}$), the estimated sample size was still lower than the required sample size when no pilot data were used (846 and 230, respectively). In contrast, for $p_C = 0.6$, the required sample size without a pilot study was 208, and incongruous pilot results (e.g., $\text{RR}_{\text{pilot}} = 0.8 \times \text{RR}_{\text{true}}$) led to a slight increase in the sample size estimate (e.g., 216), reflecting a potential penalty in settings where the pilot prior is pessimistic. These results highlight that robust MAP priors can typically reduce required sample sizes, but that the benefits are sensitive to the alignment between pilot and definitive study effects.

\section{Discussion}
\label{sec:discussion}

This paper provides practical guidance for incorporating pilot data into definitive rare disease trials using robust MAP priors. While previous studies have shown that Bayesian borrowing from historical data can reduce required sample sizes \citep{qi2022sample}, our work extends these findings by demonstrating how pilot-informed priors can also substantially shorten expected trial duration and improve the probability of completing recruitment within a fixed time frame. These operational gains are especially relevant in rare disease settings, where slow recruitment and logistical constraints can threaten trial feasibility. We found that even modest pilot studies—comprising just 10–20\% of the definitive sample size—led to meaningful reductions in expected trial duration across a range of recruitment scenarios and control event risks. Additionally, simulations showed clear improvements in the probability of reaching recruitment targets on time, reinforcing the value of early investment in pilot data.

Reusing pilot data as historical priors offers major practical and ethical advantages. In rare disease research, where each patient enrolled represents a substantial investment of time, coordination, and funding, discarding high-quality pilot data at the analysis stage wastes valuable resources. Leveraging these data in the definitive trial respects the contribution of pilot study participants and maximizes the return on early stage research infrastructure. It also reduces the need to repeat similar data collection efforts, freeing up capacity for more targeted recruitment and follow-up. By using pilot data as formal prior input, investigators can increase the feasibility of rigorous trial designs without sacrificing statistical validity—an approach aligned with regulatory guidance on the use of Bayesian methods in medical research \citep{fda2010bayesian}.

\cite{schmidli2014robust} emphasize that the success of traditional, non-robust MAP priors relies heavily on the assumption of exchangeability between historical and new data. When this assumption fails—due to differences in patient populations, outcome definitions, or trial conduct, for example—MAP priors can lead to biased estimates and inflated type I error rates. To mitigate this risk, we adopted a robust MAP prior, which combines an informative component derived from pilot data with a non-informative informative component. This mixture structure reduces sensitivity to prior-data conflict while still allowing substantial borrowing when prior and new data are consistent, maintaining type I error control even when disagreement exists between prior and new data. In our setting, however, we assume a high degree of prior-data agreement because the pilot study is assumed to be designed and conducted under the same protocol as the planned definitive trial. Specifically, both studies should use identical entry criteria, outcome definitions, and data collection procedures. Ideally, the studies should be conducted by overlapping investigator teams in similar clinical settings, minimizing the likelihood of systematic differences.

While our robust prior structure is motivated by the general framework proposed by \cite{schmidli2014robust}, our formulation differs in that the informative component is derived from a single pilot study rather than a meta-analytic model over multiple historical sources. As a result, we do not model between-trial heterogeneity explicitly via a parameter, which plays a central role in the standard MAP construction. \cite{schmidli2014robust} note that when only one historical study is available, this parameter cannot be estimated from the data and must instead be specified based on expert judgment or external empirical evidence. Our approach implicitly assumes a fixed, low level of heterogeneity. However, the inclusion of a vague prior component and the use of marginal-likelihood-based weight updating serve to relax this assumption, enabling the influence of the pilot data to adapt based on its concordance with the definitive trial data. This strategy provides a pragmatic alternative in settings where only a single, protocol-aligned source of historical data is available and a full hierarchical model is not feasible.

This study demonstrates that incorporating pilot data using robust MAP priors can substantially improve the efficiency and feasibility of definitive trials. By simulating a range of plausible recruitment rates and control event probabilities, we showed that even modest pilot studies can reduce required sample sizes, shorten expected trial durations, and increase the likelihood of reaching recruitment targets within practical time frames. These operational improvements are critical in rare diseases, where each patient enrolled represents a major investment and delays in evidence generation can limit patient access to effective therapies. Importantly, the robust MAP prior structure enables partial borrowing while maintaining protection against prior-data conflict, making it a practical and principled choice for integrating pilot data when prior and current trials are closely aligned. Future work may extend this assessment to longitudinal or time-to-event outcomes, and investigate adaptive trial frameworks where prior-data conflict assessments can be incorporated at interim analyses to dynamically adjust borrowing.
\section*{Acknowledgments}
Lara Maleyeff is supported by a CANSTAT trainee award funded by CIHR grant \#262556. Shirin Golchi acknowledges support from NSERC, Canadian Institute for Statistical Sciences (CANSSI), Fonds de recherche du Qu\'ebec - Sant\'e (FRQS) and Fonds de recherche du Qu\'ebec - Nature et technologies (FRQNT). Marie Hudson acknowledges support from the McGill Interdisciplinary Initiative in Infection and Immunity.

\newpage 
\bibliographystyle{apalike}
\bibliography{references}

\end{document}